\documentclass[epj,twocolumn]{webofc}
\usepackage[varg]{txfonts}   
\graphicspath{{fig/}}
\woctitle{Powders \& Grains 2017}
\begin{document}
\title{A microscopic theory for discontinuous shear thickening of frictional granular materials}
\author{
\firstname{Kuniyasu} \lastname{Saitoh}\inst{1}\fnsep \and
\firstname{Hisao} \lastname{Hayakawa}\inst{2}\fnsep
}
\institute{
Advanced Institute for Materials Research, Tohoku University, 2-1-1 Katahira, Aoba-ku, Sendai 980-8577, Japan
\and
Yukawa Institute for Theoretical Physics, Kyoto University, Kyoto 606-8502, Japan
}
\abstract{
We extend a recent theory for the rheology of frictionless granular materials
[K.\ Suzuki and H.\ Hayakawa, Phys.\ Rev.\ Lett.\ 2015,\ \textbf{115},\ 098001] to the case of frictional disks in two dimensions.
Employing a frictional contact model for molecular dynamics simulations, we derive difference equations of the shear stress, the granular temperature,
and the spin temperature from the generalized Green-Kubo formula, where all the terms are given by microscopic expressions.
The numerical solutions of the difference equations not only describe the flow curve, but also reproduce the hysteresis of shear stress,
which can be the signature of discontinuous shear thickening of frictional disks.}
\maketitle
\section{Introduction}
\label{sec:intro}
Understanding of the rheology of granular materials is important for industrial applications so that its microscopic descriptions have been desired in science and technology \cite{lemaitre}.
It is now known that the rheology drastically changes if the fraction exceeds the jamming point,\ i.e.\ the shear stress changes from the liquid- to solid-branches of flow curves.
Besides the strong dependence on the fraction, the friction between constituent particles also triggers the anomalous flow behavior,\ i.e.\ the \emph{discontinuous shear thickening} (DST),
where the shear stress discontinuously jumps from liquid- to solid-branches, and vice versa \cite{DST0,DST1,DST2}.
Recently, a microscopic theory for the rheology of frictionless granular materials have been developed by Suzuki and one of authors \cite{MicroTheory0},
where the critical scaling of shear viscosity and that of granular temperature below jamming were well predicted without any fitting parameters.

In this paper, we apply their theory to the case of frictional disks in two-dimension such that the DST observed in molecular dynamics (MD) simulations \cite{DST0,DST2} can be explained.
%
\section{Theory}
\label{sec:theory}
First, we describe translational motions of frictional disks by the SLLOD equation \cite{LiquidTheory1}.
We assume that the system is homogeneously sheared along the $x$-axis under the Lees-Edwards boundary condition.
Then, time derivatives of the position, $\mathbf{r}_i=(x_i,y_i)$, and \emph{peculiar momentum}, $\mathbf{p}_i=(p_{ix},p_{iy})$, of $i$-th particle ($i=1,\dots,N$) are given by
\begin{equation}
\dot{\mathbf{r}}_i = m_i^{-1}\mathbf{p}_i + \dot{\boldsymbol{\gamma}}:\mathbf{r}_i~, \hspace{5mm}
\dot{\mathbf{p}}_i = \mathbf{F}_i - \dot{\boldsymbol{\gamma}}:\mathbf{p}_i~,
\label{eq:SLLOD}
\end{equation}
respectively, where $m_i$ and $\mathbf{F}_i=(F_{ix},F_{iy})$ are the mass of $i$-th particle and the force acts on the $i$-th particle, respectively.
On the right-hand-sides of Eq.\ (\ref{eq:SLLOD}), each element of the two-rank tensor, $\dot{\boldsymbol{\gamma}}$, is defined as
$\dot{\gamma}_{\alpha\beta}=\dot{\gamma}\delta_{x\alpha}\delta_{y\beta}$ ($\alpha,\beta=x,y$) with the shear rate, $\dot{\gamma}$, such that $\dot{\boldsymbol{\gamma}}:\mathbf{r}_i=(\dot{\gamma}y_i,0)$.

Next, rotational motions of the disks are described by the Euler equation of motions,
where time derivatives of the angular position, $\boldsymbol{\varphi}_i$, and angular velocity, $\boldsymbol{\omega}_i$, of $i$-th particle are given by
\begin{equation}
\dot{\boldsymbol{\varphi}}_i = \boldsymbol{\omega}_i\times\boldsymbol{\varphi}_i~, \hspace{5mm}
I_i\dot{\boldsymbol{\omega}}_i = \mathbf{M}_i
\label{eq:Euler}~,
\end{equation}
respectively.
In Eq.\ (\ref{eq:Euler}), $I_i$ and $\mathbf{M}_i$ represent the moment of inertia and the torque acts on the $i$-th particle, respectively.
From Eqs.\ (\ref{eq:SLLOD}) and (\ref{eq:Euler}), the \emph{phase variables} are defined as $\mathbf{\Gamma}\equiv\{\mathbf{r}_i,\mathbf{p}_i,\boldsymbol{\varphi}_i,\boldsymbol{\omega}_i\}$.
\subsection{Frictional contact model}
To study the DST of frictional disks, we employ a frictional contact model \cite{dem},
where the total force acts on the $i$-th particle is divided into the normal and tangential directions as $\mathbf{F}_i = \mathbf{F}_{in} + \mathbf{F}_{it}$
(the subscripts, $n$ and $t$, are, respectively, used for the normal and the tangential components).
Here, both the normal and tangential forces are modeled by linear spring-dashpot and are further decomposed into (reversible) elastic and (irreversible) viscous parts as
$\mathbf{F}_{in}\equiv\mathbf{F}_{in}^{(\mathrm{el})} + \mathbf{F}_{in}^{(\mathrm{vis})}$ and $\mathbf{F}_{it}\equiv\mathbf{F}_{it}^{(\mathrm{el})} + \mathbf{F}_{it}^{(\mathrm{vis})}$, respectively.
For later use, we also introduce the elastic and viscous parts of the total force as $\mathbf{F}_i = \mathbf{F}_i^{(\mathrm{el})} + \mathbf{F}_i^{(\mathrm{vis})}$,
where each part is decomposed into $\mathbf{F}_i^{(\mathrm{el})}\equiv\mathbf{F}_{in}^{(\mathrm{el})}+\mathbf{F}_{it}^{(\mathrm{el})}$
and $\mathbf{F}_i^{(\mathrm{vis})}\equiv\mathbf{F}_{in}^{(\mathrm{vis})}+\mathbf{F}_{it}^{(\mathrm{vis})}$.
In the MD simulations of frictional disks, the DST is enhanced for strong friction (the friction coefficient, $\mu=2$, is used for Coulomb's friction \cite{DST0,DST2}).
Therefore, in this study, we neglect the switch from static friction, $\mathbf{F}_{it}$, to dynamical one, $\mu|\mathbf{F}_{in}|$, where all the disks in contact do not slip with each other.

In the normal force, $\mathbf{F}_{in}$, the elastic and viscous parts are given by
$\mathbf{F}_{in}^{(\mathrm{el})} = k_n\sum_{j\neq i}\Theta(\xi_{ij})\xi_{ij}\mathbf{n}_{ij}$ and
$\mathbf{F}_{in}^{(\mathrm{vis})} = -\eta_n\sum_{j\neq i}\Theta(\xi_{ij})(\dot{\mathbf{r}}_{ij}\cdot\mathbf{n}_{ij})\mathbf{n}_{ij}$,
respectively, where $k_n$ and $\eta_n$ are introduced as a spring constant and viscosity coefficient, respectively.
Here, $\Theta(\xi_{ij})$ is the Heaviside step function (which is $1$ for $\xi_{ij}>0$ and 0 otherwise),
where $\xi_{ij}\equiv R_i+R_j-r_{ij}$, $R_i$ (or $R_j$), $r_{ij}\equiv|\mathbf{r}_i-\mathbf{r}_j|$, $\mathbf{n}_{ij}\equiv(\mathbf{r}_i-\mathbf{r}_j)/r_{ij}$,
and $\dot{\mathbf{r}}_{ij}\equiv\dot{\mathbf{r}}_i-\dot{\mathbf{r}}_j$
represent an overlap between the disks ($i$ and $j$), the radius of $i$-th (or $j$-th) disk, the distance between the disks,
the unit vector parallel to the relative position, and the relative velocity, respectively.

In the tangential force, $\mathbf{F}_{it}$, the elastic and viscous parts are written as
$\mathbf{F}_{it}^{(\mathrm{el})} = -k_t\sum_{j\neq i}\Theta(\xi_{ij})\int_{\tilde{t}_{ij}}^t\mathbf{v}_{ijt}(s)ds$ and
$\mathbf{F}_{it}^{(\mathrm{vis})} = -\eta_t\sum_{j\neq i}\Theta(\xi_{ij})\mathbf{v}_{ijt}$,
respectively, where $k_t$ and $\eta_t$ are the spring constant and the viscosity coefficient, respectively,
and the disks, $i$ and $j$, get into contact at time $\tilde{t}_{ij}$.
Here, $\mathbf{v}_{ijt} = \mathbf{n}_{ij}\times(R_i\boldsymbol{\omega}_i+R_j\boldsymbol{\omega}_j) + \dot{\mathbf{r}}_{ij}-(\dot{\mathbf{r}}_{ij}\cdot\mathbf{n}_{ij})\mathbf{n}_{ij}$
represents the tangential relative velocity at the contact point \cite{dem}, which is perpendicular to the normal unit vector between the disks,\ i.e.\ $\mathbf{v}_{ijt}\cdot\mathbf{n}_{ij}=0$.
Then, the elastic part, $\mathbf{F}_{it}^{(\mathrm{el})}$, is proportional to the relative displacement, $\int_{\tilde{t}_{ij}}^t\mathbf{v}_{ijt}(s)ds$,
which depends on the memory during a duration of contact, $\tau_{ij}(t)\equiv t-\tilde{t}_{ij}$.

The total torque acts on the $i$-th particle also consists of elastic and viscous parts as $\mathbf{M}_i=\mathbf{M}_i^{(\mathrm{el})}+\mathbf{M}_i^{(\mathrm{vis})}$.
If we write the tangential forces as $\mathbf{F}_{it}^{(\mathrm{el})}=\sum_{j\neq i}\mathbf{f}_{ijt}^{(\mathrm{el})}$ and $\mathbf{F}_{it}^{(\mathrm{vis})}=\sum_{j\neq i}\mathbf{f}_{ijt}^{(\mathrm{vis})}$,
each part of the torque is given by $\mathbf{M}_i^{(\mathrm{el})}=R_i\sum_{j\neq i}\mathbf{f}_{ijt}^{(\mathrm{el})}\times\mathbf{n}_{ij}$ and $\mathbf{M}_i^{(\mathrm{vis})}=R_i\sum_{j\neq i}\mathbf{f}_{ijt}^{(\mathrm{vis})}\times\mathbf{n}_{ij}$.
\subsection{Hamiltonian and Liouville equation}
\label{sub:Hamiltonian}
In our system, the Hamiltonian is introduced as $\mathcal{H} = \sum_i(\mathbf{p}_i^2/2m_i+I_i\boldsymbol{\omega}_i^2/2) + U_n + U_t$,
where $U_n$ and $U_t$ are elastic potentials stored in the normal and tangential directions, respectively.
Because the Hamiltonian depends on time through the phase variable, $\mathbf{\Gamma}$, its time derivative is given by
$\dot{\mathcal{H}}  = \sum_i(m_i^{-1}\mathbf{p}_i\cdot\dot{\mathbf{p}}_i + I_i\boldsymbol{\omega}_i\cdot\dot{\boldsymbol{\omega}}_i
-\mathbf{F}_{in}^{(\mathrm{el})}\cdot\dot{\mathbf{r}}_i-\mathbf{F}_{it}^{(\mathrm{el})}\cdot\dot{\boldsymbol{\varphi}}_i)$,
where we have used the relations, $\mathbf{F}_{in}^{(\mathrm{el})}=-\partial U_n/\partial\mathbf{r}_i$ and $\mathbf{F}_{it}^{(\mathrm{el})}=-\partial U_t/\partial\boldsymbol{\varphi}_i$.
From Eqs.\ (\ref{eq:SLLOD}) and (\ref{eq:Euler}), the time derivative is rewritten as $\dot{\mathcal{H}} = -\dot{\gamma}S\sigma_{xy}-2\mathcal{R}+\mathcal{J}$. Here,
\begin{eqnarray}
\sigma_{xy} &\equiv& \frac{1}{S}\sum_i\left(\frac{p_{ix}p_{iy}}{m_i} + y_iF_{ix}\right)~, \label{eq:sigma_xy}\\
\mathcal{R} &\equiv&-\frac{1}{2}\sum_i\left\{\mathbf{F}_i^{(\mathrm{vis})}\cdot\dot{\mathbf{r}}_i+\mathbf{M}_i^{(\mathrm{vis})}\cdot\boldsymbol{\omega}_i\right\}~, \label{eq:R}\\
\mathcal{J} &\equiv& \sum_i\left\{\mathbf{F}_{it}^{(\mathrm{el})}\cdot\left(\dot{\mathbf{r}}_i-\boldsymbol{\omega}_i\times\boldsymbol{\varphi}_i\right)+\mathbf{M}_i^{(\mathrm{el})}\cdot\boldsymbol{\omega}_i\right\}, \label{eq:J}
\end{eqnarray}
are introduced as the shear stress
\footnote{It is also written as $\sigma_{xy}=\sum_i m_i^{-1}p_{ix}p_{iy}/S + \sum_i\sum_{j\neq i}(y_i-y_j)f_{ijx}/2S$.},
the \emph{dissipation function}, and the rate of energy production by the \emph{couple stress} \cite{rotation3}, respectively,
where $S$ represents the system area.
%
\begin{table*}[t]
\centering
\caption{Coefficients in the difference equations, Eqs.\ (\ref{eq:diff_sigma})-(\ref{eq:diff_Tt}),
where $n_{ij\alpha}$ ($\alpha=x,y$) represents the $\alpha$-component of the normal unit vector between the particles ($i$ and $j$), $\mathbf{n}_{ij}$.}
\label{tab:coefficient}
\begin{tabular}{ccl}
\hline
$a_0$ & $=$ & $2\tau_\mathrm{rel} k_n^2\sum_{\substack{i,j,k,l\\(i\neq j, k\neq l)}}\langle\Theta(\xi_{ij})\Theta(\xi_{kl})\xi_{ij}\xi_{kl}r_{ij}r_{kl} n_{ijx}n_{klx}n_{ijy}n_{kly}\rangle_{\dot{\gamma}}$ \\
$a_1$ & $=$ & $(mS)^{-1}\tau_\mathrm{rel}\sum_{\substack{i,j,k\\(i\neq j, i\neq k)}}\langle 3\eta_n^2\Theta(\xi_{ij})^2 r_{ij}^2n_{ijx}^2n_{ijy}^2
+ 3\eta_t^2\langle\Theta(\xi_{ij})^2r_{ij}^2n_{ijy}^4+\eta_n^2\Theta(\xi_{ij})\Theta(\xi_{ik})\left(n_{ijx}n_{ikx}+n_{ijy}n_{iky}\right)r_{ij}r_{ik}n_{ijx}n_{ikx}n_{ijy}n_{iky}$ \\
& & $+\eta_n\eta_t\Theta(\xi_{ij})\Theta(\xi_{ik})r_{ij}r_{ik}n_{ijx}n_{ijy}n_{iky}^2\left(n_{ijx}n_{iky}-n_{ijy}n_{ikx}\right)
+\eta_t^2\Theta(\xi_{ij})\Theta(\xi_{ik})r_{ij}r_{ik}n_{ijy}^2n_{iky}^2\left(n_{ijy}n_{iky}+n_{ijx}n_{ikx}\right)\rangle_{\dot{\gamma}}$ \\
$a_2$ & $=$ & $(mS)^{-1}\tau_\mathrm{rel}\sum_{\substack{i,j,k\\(i\neq j, i\neq k)}}\langle 6\eta_t^2\Theta(\xi_{ij})^2r_{ij}^2n_{ijy}^4
+ 2\eta_t^2\langle\Theta(\xi_{ij})\Theta(\xi_{ik})r_{ij}r_{ik}n_{ijy}^2n_{iky}^2\rangle_{\dot{\gamma}}$ \\
$b_0$ & $=$ & $2\tau_\mathrm{rel} k_n^2\sum_{\substack{i,j,k,l\\(i\neq j, k\neq l)}}\langle\Theta(\xi_{ij})\Theta(\xi_{kl})\xi_{ij}\xi_{kl}r_{ij}r_{kl} n_{ijx}n_{klx}n_{ijy}n_{kly}\rangle_{\dot{\gamma}}$ \\
$b_1$ & $=$ & $(mS)^{-1}\tau_\mathrm{rel}\sum_{\substack{i,j,k\\(i\neq j, i\neq k)}}\langle 3\eta_n^2\Theta(\xi_{ij})^2 r_{ij}^2n_{ijx}^2n_{ijy}^2
+ 3\eta_t^2\langle\Theta(\xi_{ij})^2r_{ij}^2n_{ijy}^4+\eta_n^2\Theta(\xi_{ij})\Theta(\xi_{ik})\left(n_{ijx}n_{ikx}+n_{ijy}n_{iky}\right)r_{ij}r_{ik}n_{ijx}n_{ikx}n_{ijy}n_{iky}$ \\
& & $+\eta_n\eta_t\Theta(\xi_{ij})\Theta(\xi_{ik})r_{ij}r_{ik}n_{ijx}n_{ijy}n_{iky}^2\left(n_{ijx}n_{iky}-n_{ijy}n_{ikx}\right)
+\eta_t^2\Theta(\xi_{ij})\Theta(\xi_{ik})r_{ij}r_{ik}n_{ijy}^2n_{iky}^2\left(n_{ijy}n_{iky}+n_{ijx}n_{ikx}\right)\rangle_{\dot{\gamma}}$ \\
$b_2$ & $=$ & $(mS)^{-1}\tau_\mathrm{rel}\sum_{\substack{i,j,k\\(i\neq j, i\neq k)}}\langle 6\eta_t^2\Theta(\xi_{ij})^2r_{ij}^2n_{ijy}^4
+ 2\eta_t^2\langle\Theta(\xi_{ij})\Theta(\xi_{ik})r_{ij}r_{ik}n_{ijy}^2n_{iky}^2\rangle_{\dot{\gamma}}$ \\
$c_0$ & $=$ & $2\tau_\mathrm{rel} k_n^2\sum_{\substack{i,j,k,l\\(i\neq j, k\neq l)}}\langle\Theta(\xi_{ij})\Theta(\xi_{kl})\xi_{ij}\xi_{kl}r_{ij}r_{kl} n_{ijx}n_{klx}n_{ijy}n_{kly}\rangle_{\dot{\gamma}}$ \\
$c_1$ & $=$ & $(mS)^{-1}\tau_\mathrm{rel}\sum_{\substack{i,j,k\\(i\neq j, i\neq k)}}\langle 3\eta_n^2\Theta(\xi_{ij})^2 r_{ij}^2n_{ijx}^2n_{ijy}^2
+ 3\eta_t^2\langle\Theta(\xi_{ij})^2r_{ij}^2n_{ijy}^4+\eta_n^2\Theta(\xi_{ij})\Theta(\xi_{ik})\left(n_{ijx}n_{ikx}+n_{ijy}n_{iky}\right)r_{ij}r_{ik}n_{ijx}n_{ikx}n_{ijy}n_{iky}$ \\
& & $+\eta_n\eta_t\Theta(\xi_{ij})\Theta(\xi_{ik})r_{ij}r_{ik}n_{ijx}n_{ijy}n_{iky}^2\left(n_{ijx}n_{iky}-n_{ijy}n_{ikx}\right)
+\eta_t^2\Theta(\xi_{ij})\Theta(\xi_{ik})r_{ij}r_{ik}n_{ijy}^2n_{iky}^2\left(n_{ijy}n_{iky}+n_{ijx}n_{ikx}\right)\rangle_{\dot{\gamma}}$ \\
$c_2$ & $=$ & $(mS)^{-1}\tau_\mathrm{rel}\sum_{\substack{i,j,k\\(i\neq j, i\neq k)}}\langle 6\eta_t^2\Theta(\xi_{ij})^2r_{ij}^2n_{ijy}^4
+ 2\eta_t^2\langle\Theta(\xi_{ij})\Theta(\xi_{ik})r_{ij}r_{ik}n_{ijy}^2n_{iky}^2\rangle_{\dot{\gamma}}$ \\
\hline
\end{tabular}
\end{table*}

If we introduce a \emph{dynamical variable}, $A(t)$, as a function of phase variable, $\mathbf{\Gamma}$, its time derivative is given by
\begin{equation}
\frac{d}{dt}A(t) = \dot{\boldsymbol{\Gamma}}\cdot\frac{\partial}{\partial\boldsymbol{\Gamma}}A(t)\equiv i\mathcal{L}A(t)~,
\label{eq:Liouville_eq}
\end{equation}
where $i\mathcal{L}\equiv\dot{\boldsymbol{\Gamma}}\cdot(\partial/\partial\boldsymbol{\Gamma})$ is defined as the \emph{Liouville operator} \cite{LiquidTheory1}.
From Eqs.\ (\ref{eq:SLLOD}) and (\ref{eq:Euler}), the Liouville operator is decomposed as $i\mathcal{L} = i\mathcal{L}_\mathrm{el}+i\mathcal{L}_{\dot{\gamma}}+i\mathcal{L}_\mathrm{vis}$
\footnote{Each part is explicitly written as
$i\mathcal{L}_\mathrm{el}=\sum_i\{m_i^{-1}\mathbf{p}_i\cdot(\partial/\partial\mathbf{r}_i)+\mathbf{F}_i^{(\mathrm{el})}\cdot(\partial/\partial\mathbf{p}_i)
+(\boldsymbol{\omega}_i\times\boldsymbol{\varphi}_i)\cdot(\partial/\partial\boldsymbol{\varphi}_i)+I_i^{-1}\mathbf{M}_i^{(\mathrm{el})}\cdot(\partial/\partial\boldsymbol{\omega}_i)\}$,
$i\mathcal{L}_{\dot{\gamma}}=\dot{\gamma}\sum_i(y_i\partial/\partial x_i-p_{iy}\partial/\partial p_{ix})$, and
$i\mathcal{L}_\mathrm{vis}=\sum_i\{\mathbf{F}_i^{(\mathrm{vis})}\cdot(\partial/\partial\mathbf{p}_i)+I_i^{-1}\mathbf{M}_i^{(\mathrm{vis})}\cdot(\partial/\partial\boldsymbol{\omega}_i)\}$.}.

On the other hand, the time development of $N$-body distribution function, $f(t)$, is described by the \emph{Liouville equation} as
\begin{equation}
\frac{\partial}{\partial t}f(t) = -\left(\dot{\mathbf{\Gamma}}\cdot\frac{\partial}{\partial\mathbf{\Gamma}}+\frac{\partial}{\partial\mathbf{\Gamma}}\cdot\dot{\mathbf{\Gamma}}\right)f(t)
\equiv -\left(i\mathcal{L}+\Lambda\right)f(t)~,
\label{eq:f-Liouville_eq}
\end{equation}
where $\Lambda\equiv(\partial/\partial\boldsymbol{\Gamma})\cdot\dot{\boldsymbol{\Gamma}}$ is defined as the \emph{compression factor} \cite{LiquidTheory1}.
The Liouville equation is also written as $\partial f(t)/\partial t=-i\mathcal{L}^\dagger f(t)$,
where $i\mathcal{L}^\dagger\equiv i\mathcal{L}+\Lambda$ is the adjoint of Liouville operator
so that the Liouville operator is non-Hermitian,\ i.e.\ $i\mathcal{L}^\dagger\neq i\mathcal{L}$ \cite{LiquidTheory1}.
From Eqs.\ (\ref{eq:SLLOD}) and (\ref{eq:Euler}), we find that the compression factor is given by
\begin{equation}
\Lambda = \sum_i\left\{\frac{\partial}{\partial\mathbf{p}_i}\cdot\mathbf{F}_i^{(\mathrm{vis})} + \frac{\partial}{\partial\boldsymbol{\omega}_i}\cdot(I_i^{-1}\mathbf{M}_i^{(\mathrm{vis})})\right\}
\label{eq:Lambda_vis}
\end{equation}
which only consists of irreversible viscous parts, $\mathbf{F}_i^{(\mathrm{vis})}$ and $\mathbf{M}_i^{(\mathrm{vis})}$.
Then, it is readily found that the first and second terms on the right-hand-side of Eq.\ (\ref{eq:Lambda_vis}) are reduced to
$(\partial/\partial\mathbf{p}_i)\cdot\mathbf{F}_i^{(\mathrm{vis})}=-m_i^{-1}(\eta_n+\eta_t)z_i$ and 
$(\partial/\partial\boldsymbol{\omega}_i)\cdot(I_i^{-1}\mathbf{M}_i^{(\mathrm{vis})})=-I_i^{-1}R_i^2\eta_tz_i$, respectively,
where $z_i\equiv\sum_{j\neq i}\Theta(\xi_{ij})$ is introduced as the coordination number of the $i$-th particle.
If we use the moment of inertia for a two-dimensional disk, $I_i=m_iR_i^2/2$, and assume that every mass is identical, $m_i=m$,
we find that the compression factor is proportional to the total number of contacts,\ i.e.\ $\Lambda=-(\eta_n+3\eta_t)N_c/m$ with $N_c\equiv\sum_i z_i$.
\subsection{Generalized Green-Kubo formula}
\label{sub:Green-Kubo}
To describe the shear stress of frictional disks, let us remind the measurement of \emph{flow curves} \cite{DST0}:
After the system is equilibrated under shear, the shear rate is increased from $\dot{\gamma}$ to $\dot{\gamma}+\Delta\dot{\gamma}$ (at time $t=0$)
such that the shear stress relaxes from $\langle\sigma_{xy}\rangle_{\dot{\gamma}}$ to $\langle\sigma_{xy}\rangle_{\dot{\gamma}+\Delta\dot{\gamma}}$,
where $\langle\dots\rangle_{\dot{\gamma}}$ represents the average in a non-equilibrium steady state under shear with $\dot{\gamma}$.
The \emph{generalized Green-Kubo formula} around the steady state characterized by $\dot{\gamma}$ is expressed as
\begin{equation}
\langle\sigma_{xy}(t)\rangle_{\dot{\gamma}+\Delta\dot{\gamma}}\simeq\langle\sigma_{xy}\rangle_{\dot{\gamma}} + \int_0^t ds\langle\sigma_{xy}(s)\Omega\rangle_{\dot{\gamma}}~,
\label{eq:GGK_sigma}
\end{equation}
where $\Omega\equiv\beta\dot{\mathcal{H}}-\Lambda=-\dot{\gamma}\beta S\sigma_{xy}-2\beta\mathcal{R}+\beta\mathcal{J}-\Lambda$
with the inverse of granular temperature, $\beta\equiv 1/\langle T\rangle_{\dot{\gamma}}$, is the \emph{work function} \cite{LiquidTheory1,MicroTheory3}
and $\langle\sigma_{xy}\rangle_{\dot{\gamma}}\neq 0$.
Assuming an exponential decay of the time correlation function,
$\langle\sigma_{xy}(s)\Omega\rangle_{\dot{\gamma}}\simeq\langle\sigma_{xy}\Omega\rangle_{\dot{\gamma}}e^{-s/\tau_\mathrm{rel}}$, with a relaxation time, $\tau_\mathrm{rel}$,
and taking a long time limit ($t\rightarrow\infty$), we reduce Eq.\ (\ref{eq:GGK_sigma}) to
$\langle\sigma_{xy}\rangle_{\dot{\gamma}+\Delta\dot{\gamma}}\simeq\langle\sigma_{xy}\rangle_{\dot{\gamma}} + \tau_\mathrm{rel}\langle\sigma_{xy}\Omega\rangle_{\dot{\gamma}}$,
where we have rewritten the steady state value as $\langle\sigma_{xy}\rangle_{\dot{\gamma}+\Delta\dot{\gamma}}\equiv\langle\sigma_{xy}(\infty)\rangle_{\dot{\gamma}+\Delta\dot{\gamma}}$
(see Ref.\ \cite{MicroTheory0} for more rigorous treatment).
If we assume that the non-equilibrium steady state is not far from the equilibrium state, we can rewrite the correlation function as
$\langle\sigma_{xy}\Omega\rangle_{\dot{\gamma}}\simeq\langle\sigma_{xy}\Omega\rangle_\mathrm{eq}+\tau_\mathrm{rel}\langle\sigma_{xy}\Omega^2\rangle_\mathrm{eq}$,
where $\langle\dots\rangle_\mathrm{eq}$ represents the ensemble average in the equilibrium state.
Then, we find that the shear stress is expanded into the series of relaxation time as
$\langle\sigma_{xy}\rangle_{\dot{\gamma}+\Delta\dot{\gamma}}\simeq\langle\sigma_{xy}\rangle_{\dot{\gamma}}+\tau_\mathrm{rel}\langle\sigma_{xy}\Omega\rangle_\mathrm{eq}+\tau_\mathrm{rel}^2\langle\sigma_{xy}\Omega^2\rangle_\mathrm{eq}$.
The \emph{parity} of phase variables requires $\langle\sigma_{xy}\Omega^2\rangle_\mathrm{eq}=0$ \cite{MicroTheory0}
and finally, the shear stress is given by
$\langle\sigma_{xy}\rangle_{\dot{\gamma}+\Delta\dot{\gamma}}\simeq\langle\sigma_{xy}\rangle_{\dot{\gamma}}+\tau_\mathrm{rel}\langle\sigma_{xy}\Omega\rangle_\mathrm{eq}$,
where the equilibrium averages, $\langle\sigma_{xy}\Omega\rangle_\mathrm{eq}=-\dot{\gamma}\beta S\langle\sigma_{xy}^2\rangle_\mathrm{eq}-2\beta\langle\sigma_{xy}\mathcal{R}\rangle_\mathrm{eq}+\beta\langle\sigma_{xy}\mathcal{J}\rangle_\mathrm{eq}-\langle\sigma_{xy}\Lambda\rangle_\mathrm{eq}$, can be explicitly calculated from Eqs.\ (\ref{eq:sigma_xy})-(\ref{eq:J}) and (\ref{eq:Lambda_vis}).
Applying the parallel procedure to the granular temperature, $\langle T\rangle_{\dot{\gamma}}$, and rotational temperature, $\langle T_t\rangle_{\dot{\gamma}}$
\footnote{The granular temperature and rotational temperature are introduced as $T\equiv N^{-1}\sum_i\mathbf{p}_i^2/2m$ and $T_t\equiv N^{-1}\sum_i I_i\omega_i^2$, respectively.},
we find difference equations as
\begin{eqnarray}
\langle\sigma_{xy}\rangle_{\dot{\gamma}+\Delta\dot{\gamma}} &=& \langle\sigma_{xy}\rangle_{\dot{\gamma}}
-a_0-a_1\dot{\gamma}-\frac{\dot{\gamma}h(\dot{\gamma})+a_2\dot{\gamma}\langle T_t\rangle_{\dot{\gamma}}}{\langle T\rangle_{\dot{\gamma}}}~,\nonumber\\
\label{eq:diff_sigma} \\
\langle T\rangle_{\dot{\gamma}+\Delta\dot{\gamma}} &=& (b_1+1)\langle T\rangle_{\dot{\gamma}}+b_0\dot{\gamma}^2-b_2\langle T_t\rangle_{\dot{\gamma}}~, \label{eq:diff_T} \\
\langle T_t\rangle_{\dot{\gamma}+\Delta\dot{\gamma}} &=& (c_1+1)\langle T_t\rangle_{\dot{\gamma}} + (c_0\dot{\gamma}^2-c_2\langle T_t\rangle_{\dot{\gamma}})\frac{\langle T_t\rangle_{\dot{\gamma}}}{\langle T\rangle_{\dot{\gamma}}},\nonumber\\
\label{eq:diff_Tt}
\end{eqnarray}
where the coefficients ($a_0$, $a_1$, $a_2$, $b_0$, $b_1$, $b_2$, $c_0$, $c_1$, and $c_2$) are listed in Table \ref{tab:coefficient}
\footnote{To calculate the correlation functions, the elastic part of tangential force is approximated to
$\mathbf{F}_{it}^{(\mathrm{el})}\simeq-k_t\sum_{j\neq i}\Theta(\xi_{ij})\tau_{ij}\mathbf{v}_{ijt}$.}.

To explicitly calculate the coefficients in Table \ref{tab:coefficient}, we need to calculate the three- and four-body correlations \cite{MicroTheory0}
and have to introduce the radial distribution function for frictional disks.
In addition, it is necessary to determine the two time scales,\ i.e.\ the relaxation time, $\tau_\mathrm{rel}$, and duration of contact, $\tau_{ij}$, which is far beyond the scope of this work.
Therefore, in the following analysis, we simply regard the coefficients as fitting parameters.
\subsection{Flow curves}
\begin{figure}
\centering
\includegraphics[width=0.5\textwidth,clip]{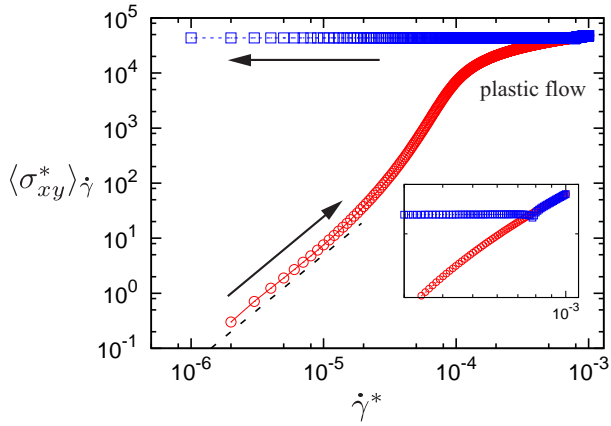}
\caption{(Color online)
Numerical results of scaled shear stress, $\langle\sigma_{xy}^\ast\rangle_{\dot{\gamma}}$.
The scaled shear rate, $\dot{\gamma}^\ast$, is first increased (the open circles) and then decreased (the open squares) as shown by the two arrows,
where the scaled increment is given by $\Delta\dot{\gamma}^\ast=10^{-6}$.
The dotted line represents Bagnold's scaling, $\langle\sigma_{xy}^\ast\rangle_{\dot{\gamma}}\sim\dot{\gamma}^{\ast 2}$, and the inset shows the zoom-in to the plastic flow regime.
Here, we choose numerical values of the coefficients, $a_0^\ast=2.0\times10^{-7}$, $a_1^\ast=4.0\times10^{-7}$, $a_2^\ast=1.0\times10^{-7}$, $b_0^\ast=3.0\times10^{-7}$, $b_1^\ast=1.6\times10^{-7}$, $b_2^\ast=1.0\times10^{-7}$,
$c_0^\ast=b_0^\ast$, $c_1^\ast=b_1^\ast$, and $c_2^\ast=b_2^\ast$.}
\label{fig:flow_curve}
\end{figure}

To draw the flow curve of frictional disks, we numerically solve the difference equations (\ref{eq:diff_sigma})-(\ref{eq:diff_Tt}).
Here, we scale the mass, length, and time by $m$, $d$ (the diameter of disk), and $t_0\equiv\eta_n/k_n$
such that the shear stress, granular (rotational) temperature, and shear rate are scaled as $\langle\sigma_{xy}^\ast\rangle_{\dot{\gamma}}\equiv(t_0^2/m)\langle\sigma_{xy}\rangle_{\dot{\gamma}}$,
$\langle T^\ast\rangle_{\dot{\gamma}}\equiv(t_0^2/md^2)\langle T\rangle_{\dot{\gamma}}$ ($\langle T_t^\ast\rangle_{\dot{\gamma}}\equiv(t_0^2/md^2)\langle T_t\rangle_{\dot{\gamma}}$),
and $\dot{\gamma}^\ast\equiv t_0\dot{\gamma}$, respectively.
In addition, the coefficients on the right-hand-sides of Eqs.\ (\ref{eq:diff_sigma})-(\ref{eq:diff_Tt}) are nondimensionalized as
$a_0^\ast\equiv(t_0^3/m)a_0$, $a_1^\ast\equiv(t_0^2/m)a_1$, $a_2^\ast\equiv(t_0^2/m)a_2$, $b_0^\ast\equiv(t_0/md^2)b_0$, $b_1^\ast\equiv t_0b_1$, $b_2^\ast\equiv t_0b_2$,
$c_0^\ast\equiv(t_0/md^2)c_0$, $c_1^\ast\equiv t_0c_1$, and $c_2^\ast\equiv t_0c_2$, which are used as numerical parameters (see the caption of Fig.\ \ref{fig:flow_curve}).

Figure \ref{fig:flow_curve} displays numerical solutions of scaled shear stress, where the shear rate increases (the open circles) and then decreases (the open squares) as shown by the two arrows.
Increasing the shear rate, we find that the shear stress satisfies \emph{Bagnold's scaling},\ i.e.\ $\langle\sigma_{xy}^\ast\rangle_{\dot{\gamma}}\sim\dot{\gamma}^{\ast 2}$ (the dotted line)
in the slow flow regime ($\dot{\gamma}^\ast\lesssim10^{-5}$), while it becomes linear,\ i.e.\ $\langle\sigma_{xy}^\ast\rangle_{\dot{\gamma}}\sim\dot{\gamma}^\ast$ (the inset)
in the plastic flow regime ($\dot{\gamma}^\ast\gtrsim10^{-4}$) as known by the \emph{$\mu$-$I$ rheology} \cite{muI0}.
On the other hand, if we decrease the shear rate, the shear stress exhibits irreversibility, where it becomes almost rate-independent,\
i.e.\ $\langle\sigma_{xy}^\ast\rangle_{\dot{\gamma}}\sim\sigma_Y$ with the \emph{yield stress}, $\sigma_Y$, in the solid-like regime ($\dot{\gamma}^\ast\lesssim10^{-4}$).
Therefore, our model, Eqs.\ (\ref{eq:diff_sigma})-(\ref{eq:diff_Tt}), well captures not only all the flow curve of frictional disks,
but also the \emph{hysteresis} of shear stress, which is the signature of the DST.
\section{Summary}
\label{sec:disc}
In this study, we have extended the previous microscopic theory for frictionless granular particles \cite{MicroTheory0} to the case of two-dimensional frictional disks,
where all the flow behavior,\ i.e.\ Bagnold's scaling, $\langle\sigma_{xy}^\ast\rangle_{\dot{\gamma}}\sim\dot{\gamma}^{\ast 2}$,
the plastic flow, $\langle\sigma_{xy}^\ast\rangle_{\dot{\gamma}}\sim\dot{\gamma}^\ast$, the rate-independent yielding stress, $\langle\sigma_{xy}^\ast\rangle_{\dot{\gamma}}\sim\sigma_Y$,
and the hysteresis of shear stress as the signature of DST, is well described by the difference equations (\ref{eq:diff_sigma})-(\ref{eq:diff_Tt}).
All the terms in Eqs.\ (\ref{eq:diff_sigma})-(\ref{eq:diff_Tt}) including the coefficients have been derived from the generalized Green-Kubo formula,\
Eq.\ (\ref{eq:GGK_sigma}), where the canonical ensembles are assumed for the system under shear and the time correlation function between the shear stress and work function,\ i.e.\
$\langle\sigma_{xy}(s)\Omega\rangle_{\dot{\gamma}}$, is approximated by the exponential function with the relaxation time, $\tau_\mathrm{rel}$.
Although it is needed to introduce the radial distribution function for frictional disks and determine the two time scales,\ i.e.\ the relaxation time and duration of contact,
the coefficients in the difference equations have been provided by the microscopic expressions as listed in Table \ref{tab:coefficient}.
However, we have used the coefficients as numerical parameters to draw the flow curves of frictional disks (Fig.\ \ref{fig:flow_curve}).
From our results, we can expect that the coupling with rotational temperature, which is absent in the case of frictionless particles \cite{MicroTheory0},
triggers both the transition from the liquid-like branch to solid-like one and the hysteresis behavior of shear stress.

Nevertheless, the transition from the liquid-like behavior (unjammed state) to the solid-like branch (jammed state)
during the increase of shear rate (the open circles in Fig.\ \ref{fig:flow_curve}) is much smoother than that observed in MD simulations \cite{DST0,DST2}.
In addition, we have not observed a sudden drop from the solid-like branch to the liquid-like one (the open squares in Fig.\ \ref{fig:flow_curve}).
Therefore, we will need further investigations of our microscopic approach to the DST, where more rigorous treatments of the coefficients, time scales, and the radial distribution function might be crucial.
In addition, it is important to implement the recent theory of non-Brownian suspensions,
where the competing divergences of shear viscosity with the smooth shift of jamming point explain the discontinuous shear thickening \cite{DST3}.

In conclusion, we have developed a microscopic theory for the DST of two-dimensional frictional disks,
where the difference equations of the shear stress, the granular temperature, and the rotational temperature
derived from the generalized Green-Kubo formula well explain all the flow behavior including the hysteresis of the shear stress.
\section*{Acknowledgements}
We thank K.\ Suzuki, S.\ Takada, M.\ Otsuki, and T.\ Kawasaki for fruitful discussions.
This work was supported by the World Premier International Research Center Initiative (WPI), Ministry of Education, Culture, Sports, Science, and Technology, Japan (MEXT),
Kawai Foundation for Sound Technology \& Music, Grant-in-Aid for Scientific Research B (Grants No. 16H04025),
and Grant-in-Aid for Research Activity Start-up (Grant No. 16H06628) from the Japan Society for the Promotion of Science (JSPS).
%
\bibliography{yield}

\begin{thebibliography}{11}

\bibitem{lemaitre}
J.~Lemaitre, J.L. Chaboche, \emph{Mechanics of Solid Materials} (Cambridge
  University Press, Cambridge, UK, 1990)

\bibitem{DST0}
M.~Otsuki, H.~Hayakawa, Phys.\ Rev.\ E \textbf{83}, 051301 (2011)

\bibitem{DST1}
R.~Seto, R.~Mari, J.F. Morris, M.M. Denn, Phys.\ Rev.\ Lett. \textbf{111},
  218301 (2013)

\bibitem{DST2}
M.~Grob, C.~Heussinger, A.~Zippelius, Phys.\ Rev.\ E \textbf{89}, 050201(R)
  (2014)

\bibitem{MicroTheory0}
K.~Suzuki, H.~Hayakawa, Phys.\ Rev.\ Lett. \textbf{115}, 098001 (2015)

\bibitem{LiquidTheory1}
D.J. Evans, G.~Morriss, \emph{Statistical Mechanics of Nonequilibrium Liquids},
  2nd~edn. (Cambridge University Press, Cambridge, UK, 2008)

\bibitem{dem}
S.~Luding, J.\ Phys.:\ Condens.\ Matter \textbf{17}, S2623 (2005)

\bibitem{rotation3}
N.~Mitarai, H.~Hayakawa, H.~Nakanishi, Phys.\ Rev.\ Lett. \textbf{88}, 174301
  (2002)

\bibitem{MicroTheory3}
H.~Hayakawa, M.~Otsuki, Phys.\ Rev.\ E \textbf{88}, 032117 (2013)

\bibitem{muI0}
G.D.R. MiDi, Eur. Phys. J. E \textbf{14}, 341 (2004)

\bibitem{DST3}
M.~Wyart, M.E. Cates, Phys.\ Rev.\ Lett. \textbf{112}, 098302 (2014)

\end{thebibliography}
%
%
%
%
\end{document}